\documentclass[12pt]{article}

\usepackage{algorithm}				% algorithms
\usepackage[noend]{algpseudocode}	% algorithms
\usepackage{amsmath}				% math package
\usepackage{amssymb}				% math package
\usepackage{amsthm}				% math package
\usepackage{caption}
\usepackage{subcaption}
\usepackage{color}
\usepackage[margin=1in]{geometry}
\usepackage[hidelinks]{hyperref}
\usepackage{titling}

\usepackage{tikz}
\usetikzlibrary{matrix}

\usepackage{savesym}			% workaround to make dingbat work with amsfonts
\savesymbol{checkmark}
\usepackage{dingbat}			% carriage return symbol

\theoremstyle{plain}				% AMS theorem styles
\newtheorem{theorem}{Theorem}
\newtheorem{lemma}[theorem]{Lemma}
\newtheorem{proposition}[theorem]{Proposition}

\theoremstyle{definition}			% AMS definition styles
\newtheorem{definition}[theorem]{Definition}

\theoremstyle{remark}			% AMS remark styles
\newtheorem*{remark}{Remark}

\newcommand{\TDFA}{\textsf{2DFA-4W}}

\newcommand{\TDFATW}{\textsf{2DFA-3W}}

\newcommand{\TDFATWOW}{\textsf{2DFA-2W}}

\newcommand{\TNFA}{\textsf{2NFA-4W}}

\newcommand{\TNFATW}{\textsf{2NFA-3W}}

\newcommand{\TNFATWOW}{\textsf{2NFA-2W}}

\newcommand{\BFA}{\textsf{BFA}}
\newcommand{\BDFA}{\textsf{BDFA}}
\newcommand{\RFA}{\textsf{RFA}}
\newcommand{\TYFA}{\textsf{TFA}}
\newcommand{\TYDFA}{\textsf{TDFA}}

\renewcommand{\phi}{\varphi}

\thanksmarkseries{alph}

\title{Two-Dimensional Typewriter Automata}
\author{
	Taylor J. Smith \thanks{Department of Computer Science, St.\ Francis Xavier University, Antigonish, Nova Scotia, Canada. Email: \href{mailto:tjsmith@stfx.ca}{\texttt{tjsmith@stfx.ca}}.}
}
\date{\today}

\begin{document}

%%%%%%

\maketitle

\begin{abstract}
A typewriter automaton is a special variant of a two-dimensional automaton that receives two-dimensional words as input and is only capable of moving its input head through its input word in three directions: downward, leftward, and rightward. In addition, downward and leftward moves may only be made via a special ``reset" move that simulates the action of a typewriter's carriage return.

In this paper, we initiate the study of the typewriter automaton model and relate it to similar models, including three-way two-dimensional automata, boustrophedon automata, and returning automata. We study the recognition powers of the typewriter automaton model, establish closure properties of the class of languages recognized by the model, and consider operational state complexity bounds for the specific operation of row concatenation. We also provide a variety of potential future research directions pertaining to the model.

\medskip


\medskip

\noindent\textit{MSC2020 classes:} 68Q45 (primary); 68Q15, 68Q19 (secondary).
\end{abstract}

%%%%%%

\section{Introduction}\label{sec:introduction}

A two-dimensional typewriter automaton is a variant of the three-way two-dimensional automaton model introduced by Rosenfeld~\cite{Rosenfeld1979PictureLanguages}. Similar to the three-way model, the input head of a typewriter automaton can move downward, leftward, and rightward, but the downward and leftward moves work in a fashion similar to the carriage of a typewriter: a special ``return" transition repeatedly moves the input head leftward to the leftmost symbol of the current row, then downward by one row. Thus, unlike the three-way two-dimensional automaton model, a typewriter automaton cannot make individual downward or leftward moves.

The typewriter automaton is similar to two other unconventional two-dimensional automaton models that have appeared in the literature~\cite{Fernau2015ScanningPicturesBoustrophedon}: the boustrophedon automaton, which moves its input head from left to right and from right to left on alternating rows of its input word; and the returning automaton, which moves its input head from the leftmost symbol to the rightmost symbol of each row of its input word.

In this paper, we initiate the study of typewriter automata by considering the recognition and closure properties of both deterministic and nondeterministic variants of the model. We show that typewriter automata are capable of recognizing a class of languages ``in between" the classes of languages recognized by two-way and three-way two-dimensional automata, and that the class of languages recognized by nondeterministic typewriter automata is equivalent to the class of languages recognized by boustrophedon automata. We further show that the class of languages recognized by typewriter automata is closed under a variety of language operations, and we establish an operational state complexity bound for the specific operation of row concatenation. We conclude by offering a number of potential directions for future research on this model.

%%%%%%

\section{Preliminaries}\label{sec:preliminaries}

A two-dimensional word consists of a finite array, or rectangle, of cells each labelled by a symbol from a finite alphabet $\Sigma$. We denote the set of $m \times n$ two-dimensional words over $\Sigma$ by $\Sigma^{m \times n}$. When a two-dimensional word is written on the input tape of a two-dimensional automaton, the cells around the word are labelled by a special boundary marker $\# \not\in \Sigma$. A two-dimensional automaton has a finite state control that is capable of moving its input head in four directions within its input word: up, down, left, and right (denoted by $U$, $D$, $L$, and $R$, respectively).

\begin{definition}[Two-dimensional automaton]\label{def:2DFA}
A two-dimensional automaton is a tuple \allowbreak $(Q, \Sigma, \delta, q_{0}, q_{\rm accept})$, where $Q$ is a finite set of states, $\Sigma$ is the input alphabet (with $\# \not\in \Sigma$ acting as a boundary marker), $\delta: (Q \setminus \{q_{\rm accept}\}) \times (\Sigma \cup \{\#\}) \to Q \times \{U, D, L, R\}$ is the partial transition function, and $q_{0}, q_{\rm accept} \in Q$ are the initial and accepting states, respectively.
\end{definition}

We can specify a nondeterministic variant of the model given in Definition~\ref{def:2DFA} by changing the transition function to map to $2^{Q \times \{U, D, L, R\}}$. We denote the deterministic and nondeterministic two-dimensional automaton models by \TDFA\ and \TNFA, respectively.

By restricting the movement of the input head to prohibit upward movements, we obtain the restricted three-way variant of the two-dimensional automaton model.

\begin{definition}[Three-way two-dimensional automaton]\label{def:2DFA3W}
A three-way two-dimensional automaton is a tuple $(Q, \Sigma, \delta, q_{0}, q_{\rm accept})$ as in Definition~\ref{def:2DFA}, where the transition function $\delta$ is restricted to use only the directions $\{D, L, R\}$.
\end{definition}

We denote deterministic and nondeterministic three-way two-dimensional automata by \TDFATW\ and \TNFATW, respectively. Likewise, we may restrict both upward and leftward movements to obtain the two-way two-dimensional automaton model, denoted in the deterministic and nondeterministic cases by \TDFATWOW\ and \TNFATWOW, respectively.

Additional information about the two-dimensional automaton model can be found in surveys by Inoue and Takanami~\cite{Inoue19912DAutomataSurvey}, Kari and Salo~\cite{KariSalo2011PictureWalkingAutomataSurvey}, and the author~\cite{Smith2019TwoDimensionalAutomata}.

Related to the two-dimensional automaton model is the notion of a boustrophedon automaton, so named because the input head of such an automaton moves from left to right and from right to left on alternating rows of its input word, as in the boustrophedon style of writing. Although the name ``boustrophedon automaton" was originally used by P\'{e}cuchet in the 1980s in reference to the two-way one-dimensional (string) automaton model~\cite{Pecuchet1985BoustrophedonInfinite, Pecuchet1985BoustrophedonBirget}, the boustrophedon automaton model considered in this paper was introduced by Fernau et al.\ in 2015~\cite{Fernau2015ScanningPicturesBoustrophedon} and studied further in subsequent papers~\cite{Fernau2018SimplePictureProcessing, Fernau2016RegularArrayBoustrophedon}.

\begin{definition}[Boustrophedon automaton]
A (two-dimensional) boustrophedon automaton is a tuple $(Q, \Sigma, R, q_{0}, q_{\text{accept}}, \#, \square)$, where $Q$, $\Sigma$, $q_{0}$, and $q_{\text{accept}}$ are as in Definition~\ref{def:2DFA}, $R \subseteq Q \times (\Sigma \cup \{\#\}) \times Q$ is a finite set of rules, $\# \not\in \Sigma$ is a special boundary marker, and $\square$ indicates an erased portion of the input word.
\end{definition}

The set of rules $R$ corresponding to a given boustrophedon automaton act as the transition relation of the automaton. 
The special marker $\square$, which indicates an erased portion of the input word, is used to distinguish symbols that have previously been read by the input head in a given configuration of an automaton. The use of the special marker $\square$ is vital to defining the acceptance criterion for a boustrophedon automaton; since such an automaton must read all of the symbols in its input word before accepting, 
an $m \times n$ word is accepted if there exists some sequence of configurations ending in an accepting state with $m \times n$ special markers $\square$ on the input tape.
A complete description of the boustrophedon automaton model can be found in the papers by Fernau et al.~\cite{Fernau2015ScanningPicturesBoustrophedon, Fernau2018SimplePictureProcessing, Fernau2016RegularArrayBoustrophedon}. We denote boustrophedon automata and their deterministic variant by \BFA\ and \BDFA, respectively.

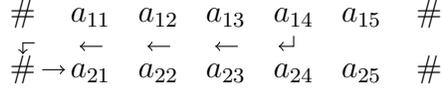
\begin{figure}
\centering
\begin{tikzpicture}
\matrix[matrix of nodes,nodes={inner sep=0pt,text width=.9cm,align=center,minimum height=.7cm}]{
\node(a01){\#};	& \node(a11){$a_{11}$};	& \node(a21){$a_{12}$};	& \node(a31){$a_{13}$};	& \node(a41){$a_{14}$};	& \node(a51){$a_{15}$};	& \# \\
\node(a02){\#};	& \node(a12){$a_{21}$};	& \node(a22){$a_{22}$};	& \node(a32){$a_{23}$};	& \node(a42){$a_{24}$};	& \node(a52){$a_{25}$};	& \# \\
};

\draw[->] (a41.center)+(0mm,-2.5mm) -- (a42.north) -- +(-2mm,0mm);
\draw[->] (a32.north east)+(-3mm,0mm) -- +(-6mm,0mm);
\draw[->] (a22.north east)+(-3mm,0mm) -- +(-6mm,0mm);
\draw[->] (a12.north east)+(-3mm,0mm) -- +(-6mm,0mm);
\draw[->] (a02.north east)+(-3mm,0mm) -- (a02.north) -- +(0mm,-1.5mm);
\draw[->] (a02.east)+(-2mm,0mm) -- +(1mm,0mm);
\end{tikzpicture}
\caption{An illustration of a typewriter automaton making a reset move \carriagereturn\ after reading a symbol $a_{14}$. Each of the moves indicated by arrows is made instantaneously; that is, symbols $a_{13}$, $a_{12}$, and $a_{11}$ are not read again as the input head moves to symbol $a_{21}$.}
\label{fig:resetmove}
\end{figure}

We now turn to the model that forms the basis of this paper: the typewriter automaton. Like a three-way two-dimensional automaton, a typewriter automaton is only capable of moving its input head in three directions: downward, leftward, and rightward. At the same time, like a boustrophedon automaton, a typewriter automaton reads its input word row-by-row, with the distinction that it moves its input head only from left to right upon entering a given row of its input word. The name of the model is inspired by its behaviour when moving between rows: a typewriter automaton makes use of a special ``reset" move, denoted by \carriagereturn\ and depicted in Figure~\ref{fig:resetmove}, which simulates the action of a typewriter's carriage return.

\begin{definition}[Typewriter automaton]
A (two-dimensional) typewriter automaton is a tuple $(Q, \Sigma, \delta, q_{0}, q_{\rm accept})$, where $Q$, $\Sigma$, $q_{0}$, and $q_{\text{accept}}$ are as in Definition~\ref{def:2DFA}, and where the transition function $\delta$ is restricted to use only the directions $\{R, \text{\carriagereturn}\}$. The reset move \carriagereturn\ shifts the input head leftward until it reaches the leftmost boundary marker of the current row, then moves the input head one position downward and one position rightward to the first symbol of the subsequent row.
\end{definition}

We denote typewriter automata and their deterministic variant by \TYFA\ and \TYDFA, respectively.

Although the typewriter automaton model is quite similar to the three-way two-dimensional automaton model, one notable difference between the two models is that the typewriter automaton cannot make individual downward or leftward input head moves. Once the reset move \carriagereturn\ is used, the input head moves repeatedly leftward and once downward without reading the symbol underneath. The input head only commences reading symbols again once its input head is positioned over the first symbol of the following row.

We define the notions of a configuration of a typewriter automaton and an accepting computation of a typewriter automaton identically to those used for traditional two-dimensional automata~\cite{Smith2019TwoDimensionalAutomata}. A configuration of a typewriter automaton $\mathcal{T}$ on an input word $w \in \Sigma^{m \times n}$ is a tuple $(q, i, j)$, where $q$ is the current state of $\mathcal{T}$ and $0 \leq i \leq m+1$ and $0 \leq j \leq n+1$ are the current positions of the input head in $w$. We then say that $\mathcal{T}$ accepts its input word $w$ if, at some point during its computation, it enters a configuration $(q, i, j)$ where $q = q_{\text{accept}}$.

To simplify the acceptance criterion for typewriter automata, we will not assume that all symbols of the input word must be read. While this choice differs from the definition of acceptance used by boustrophedon automata, it is more in line with the definition used for other two-dimensional automaton models.

\begin{remark}
Fernau et al.\ also studied the so-called returning automaton model, which operates in a manner similar to a boustrophedon automaton, but always moves its input head from the leftmost symbol to the rightmost symbol in a given row~\cite{Fernau2015ScanningPicturesBoustrophedon}. In this sense, a returning automaton is syntactically identical to a boustrophedon automaton, but distinct from a typewriter automaton since typewriter automata are not required to read entire rows of their input words.
\end{remark}

%%%%%%

\section{Recognition Properties}\label{sec:recognition}

It is well-known that three-way two-dimensional automata recognize strictly fewer languages than the standard (four-way) two-dimensional automaton model~\cite{Rosenfeld1979PictureLanguages}, and a similar relationship holds between two- and three-way two-dimensional automata. Likewise, all deterministic 
two-dimensional automaton models recognize strictly fewer languages than their nondeterministic counterparts, resulting in a lattice-like recognition hierarchy between deterministic and nondeterministic two-, three-, and four-way two-dimensional automata~\cite{Rosenfeld1979PictureLanguages}.

In their original paper, Fernau et al.\ showed that both deterministic and nondeterministic boustrophedon automata recognize the same class of two-dimensional languages, as they process their input in a manner analogous to a one-dimensional (string) automaton~\cite{Fernau2015ScanningPicturesBoustrophedon}. Moreover, since boustrophedon automata and returning automata are syntactically equivalent, Fernau et al.\ further showed that both of these models recognize the same class of two-dimensional languages~\cite{Fernau2015ScanningPicturesBoustrophedon}. However, despite their similarities in input head movement, the class of languages recognized by boustrophedon automata is a strict subset of the class of languages recognized by nondeterministic three-way two-dimensional automata~\cite{Fernau2018SimplePictureProcessing}.

To see where typewriter automata are positioned in the two-dimensional automaton recognition hierarchy, we first compare the model to the weaker two-way two-dimensional automaton. This restricted variant model is similar to the three-way two-dimensional automaton in Definition~\ref{def:2DFA3W}, but it may only move its input head downward and rightward.

\begin{theorem}\label{thm:typewriter2W}
Let $L_{\text{\texttt{1}s}}$ denote the language of $2 \times 2$ words over the alphabet $\Sigma = \{\texttt{0}, \texttt{1}\}$ having \texttt{1}s at positions $(1,0)$ and $(0,1)$. Then $L_{\text{\texttt{1}s}}$ can be recognized by a typewriter automaton $\mathcal{T}_{\text{\texttt{1}s}}$, but not by any two-way two-dimensional automaton.

\begin{proof}
The typewriter automaton $\mathcal{T}_{\text{\texttt{1}s}}$ recognizes the language $L_{\text{\texttt{1}s}}$ in the following way. Starting at the top-left symbol of the input word, $\mathcal{T}_{\text{\texttt{1}s}}$ moves one position rightward to read a \texttt{1} in position $(1,0)$. It then makes a reset move to position $(0,1)$, and reads another \texttt{1} before accepting.

No two-way two-dimensional automaton can recognize the language $L_{\text{\texttt{1}s}}$, since if such an automaton recognized the language, the same accepting computation must also accept one of the $2 \times 2$ words having a \texttt{0} in either of the positions $(1,0)$ or $(0,1)$.
\end{proof}
\end{theorem}

As the previous argument works in both the deterministic and nondeterministic cases, we have that $L_{\TDFATWOW} \subset L_{\TYDFA}$ and $L_{\TNFATWOW} \subset L_{\TYFA}$.

Given our previous observation that typewriter automata and three-way two-dimensional automata are closely related, Theorem~\ref{thm:typewriter2W} may not come as much of a surprise; indeed, we previously noted that $L_{\TDFATWOW} \subset L_{\TDFATW}$ and $L_{\TNFATWOW} \subset L_{\TNFATW}$. However, the fact that typewriter automata cannot make independent downward or leftward moves is limiting enough that we can refine the relationships between all three models to show that typewriter automata are positioned between the two- and three-way two-dimensional models.

\begin{theorem}
Let $L_{\text{stairs}}$ denote the language of $n \times n$ two-dimensional words over the alphabet $\Sigma = \{\texttt{0}, \texttt{1}\}$ having a contiguous\footnote{By contiguous, we mean that pairs of \texttt{1}s are adjacent either horizontally or vertically, but not diagonally.} path of \texttt{1}s forming a staircase pattern from the top-left corner to the bottom-right corner. Then $L_{\text{stairs}}$ can be recognized by a three-way two-dimensional automaton $\mathcal{A}$, but not by any typewriter automaton.

\begin{proof}
The three-way two-dimensional automaton $\mathcal{A}$ recognizes the language $L_{\text{stairs}}$ in the following way. Starting at the top-left symbol of the input word, $\mathcal{A}$ makes an alternating series of rightward and downward moves, confirming after each move that it has read a \texttt{1}. If, at some point during the computation, $\mathcal{A}$ reads a boundary marker \# after a rightward move, it makes 
one leftward move and one downward move 
to confirm that it has read the following ``bottom-right corner" pattern:
\begin{equation*}
\begin{array}{c c}
\texttt{1}	& \# \\
 \#		& \#
\end{array}
\end{equation*}
If $\mathcal{A}$ confirms that it is in the bottom-right corner, then it accepts.

No typewriter automaton can recognize the language $L_{\text{stairs}}$, since there is no general way for such an automaton to count rightward moves and verify that the corner-to-corner path is contiguous upon making a reset move. Thus, a typewriter automaton that verifies ``treads" of the staircase pattern---i.e., pairs of horizontally adjacent \texttt{1}s in each row---could not distinguish between the subwords
\begin{equation*}
\begin{array}{c c c c c}
\#	& \#		& \#		& \#		& \# \\
\#	& \texttt{1}	& \texttt{1}	& \texttt{0}	& \texttt{0} \\
\#	& \texttt{0}	& \texttt{1}	& \texttt{1}	& \texttt{0}
\end{array}
\ \text{and} \
\begin{array}{c c c c c}
\#		& \#		& \#		& \#		& \# \\
\#		& \texttt{1}	& \texttt{1}	& \texttt{0}	& \texttt{0} \\
\#		& \texttt{0}	& \texttt{0}	& \texttt{1}	& \texttt{1}
\end{array}.
\qedhere
\end{equation*}
\end{proof}
\end{theorem}

Again, as the previous argument works in both the deterministic and nondeterministic cases, we have that $L_{\TYDFA} \subset L_{\TDFATW}$ and $L_{\TYFA} \subset L_{\TNFATW}$.

Similar to the more standard two-dimensional automaton models, we can show that a separation exists between the deterministic and nondeterministic variants of the typewriter automaton model with another small example language.

\begin{theorem}\label{thm:typewriterdeterministicnondeterministic}
Let $L_{\text{L}}$ denote the language of $m \times n$ two-dimensional words over the alphabet $\Sigma = \{\texttt{0}, \texttt{1}\}$ 
where $m, n \geq 2$, the first column and last row of each word consist entirely of \texttt{1}s, 
and all other symbols are \texttt{0}. 
Then $L_{\text{L}}$ can be recognized by a nondeterministic typewriter automaton $\mathcal{T}_{\text{L}}$, but not by any deterministic typewriter automaton.

\begin{proof}
The nondeterministic typewriter automaton $\mathcal{T}_{\text{L}}$ recognizes the language $L_{\text{L}}$ in the following way. Starting at the top-left symbol of the input word, 
$\mathcal{T}_{\text{L}}$ verifies that the symbol in the first position of the current row is a \texttt{1}, and it then makes rightward moves to check that the remainder of the row consists of \texttt{0}s. Upon reading the boundary marker \# at the end of the row, the automaton makes a reset move. 
During the computation of $\mathcal{T}_{\text{L}}$, it can nondeterministically 
guess that its input head has reached the bottom row of the input word. At this stage, the automaton reads each symbol in that row to verify whether it has read a \texttt{1}. When $\mathcal{T}_{\text{L}}$ reaches the boundary marker \#, it makes one more reset move. If, after this move, the input head reads another boundary marker \#, then $\mathcal{T}_{\text{L}}$ accepts.

No deterministic typewriter automaton can recognize the language $L_{\text{L}}$, since such an automaton 
cannot guess that its input head is in the last row of the input word. Thus, if a deterministic automaton (correctly) rejects upon reading a \texttt{1} outside of the first column and last row of the input word, then it must also reject upon reading the occurrence of \texttt{1} in the last row and second column of the same word.
\end{proof}
\end{theorem}

It is known that a nondeterministic boustrophedon automaton can recognize the same language $L_{\text{L}}$ defined in Theorem~\ref{thm:typewriterdeterministicnondeterministic} and, therefore, so too can a deterministic boustrophedon automaton~\cite{Fernau2015ScanningPicturesBoustrophedon}. More generally, given the similarities between the input head movement of a typewriter automaton and that of a returning automaton, and the equivalence in recognition power between returning automata and boustrophedon automata, we can establish the following result.

\begin{theorem}\label{thm:typewriterboustrophedonequivalent}
Nondeterministic typewriter automata and boustrophedon automata recognize the same class of languages.

\begin{proof}
To prove this result, we will show that a nondeterministic typewriter automaton can simulate the computation of a returning automaton and vice versa.

In one direction, a nondeterministic typewriter automaton $\mathcal{T}$ can recognize a language in $L_{\RFA}$ by limiting any reset moves made by the transition function of $\mathcal{T}$ to occur only after reading a boundary marker. In this way, $\mathcal{T}$ reads each row of its input word completely from left to right.

In the other direction, to allow a returning automaton $\mathcal{R}$ to recognize a language in $L_{\TYFA}$, we use the following construction. The automaton $\mathcal{R}$ begins by simulating the computation of the corresponding typewriter automaton. If, during this simulation, the typewriter automaton would have made a reset move, then $\mathcal{R}$ continues moving rightward while ignoring any non-boundary markers being read by its input head. After a boundary marker is read, $\mathcal{R}$ makes a deferred reset move and begins simulating the computation of the typewriter automaton on the next row.

Since boustrophedon automata and returning automata recognize the same class of languages, the preceding constructions show that boustrophedon automata and nondeterministic typewriter automata also recognize the same class of languages.
\end{proof}
\end{theorem}

By combining Theorem~\ref{thm:typewriterboustrophedonequivalent} with the observation that $L_{\TYDFA} \subset L_{\TYFA}$, we see that boustrophedon automata can recognize all languages recognized by deterministic typewriter automata. However, the deterministic version of Theorem~\ref{thm:typewriterboustrophedonequivalent} does not hold, since (for example) the language $L_{\text{L}}$ is recognized by a boustrophedon automaton but not by a deterministic typewriter automaton.

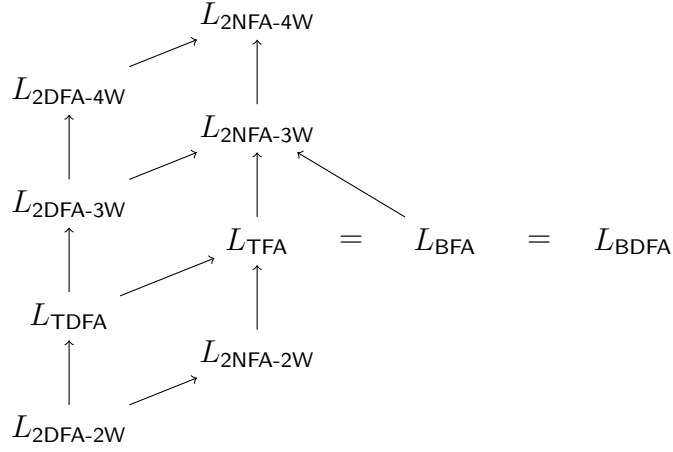
\begin{figure}
\centering
\begin{tikzpicture}
	\node (2DFA) at (0,1.5) {$L_{\TDFA}$};
	\node (2DFA-3W) at (0,0) {$L_{\TDFATW}$};
	\node (TDFA) at (0,-1.5) {$L_{\TYDFA}$};
	\node (2DFA-2W) at (0,-3) {$L_{\TDFATWOW}$};
	\node (2NFA) at (2.5,2.5) {$L_{\TNFA}$};
	\node (2NFA-3W) at (2.5,1) {$L_{\TNFATW}$};
	\node (TFA) at (2.5,-0.5) {$L_{\TYFA}$};
	\node (2NFA-2W) at (2.5,-2) {$L_{\TNFATWOW}$};
	
	\node (EQ1) at (3.75,-0.5) {$=$};
	
	\node (BFA) at (5,-0.5) {$L_{\BFA}$};
	\node (EQ2) at (6.25,-0.5) {$=$};
	\node (BDFA) at (7.5,-0.5) {$L_{\BDFA}$};
	
	\draw[->] (2DFA-2W) -- (TDFA);
	\draw[->] (TDFA) -- (2DFA-3W);
	\draw[->] (2DFA-3W) -- (2DFA);
	\draw[->] (2NFA-2W) -- (TFA);
	\draw[->] (TFA) -- (2NFA-3W);
	\draw[->] (2NFA-3W) -- (2NFA);
	\draw[->] (2DFA-2W) -- (2NFA-2W);
	\draw[->] (TDFA) -- (TFA);
	\draw[->] (2DFA-3W) -- (2NFA-3W);
	\draw[->] (2DFA) -- (2NFA);
	
	\draw[->] (BFA) -- (2NFA-3W);
\end{tikzpicture}
\caption{Inclusions among classes of languages recognized by two-dimensional automaton models mentioned in this paper. An arrow $L_{\textsf{A}} \rightarrow L_{\textsf{B}}$ indicates the relation $L_{\textsf{A}} \subset L_{\textsf{B}}$.}
\label{fig:2Dautomatainclusions}
\end{figure}

Altogether, the results established in this section produce the inclusion hierarchy depicted in Figure~\ref{fig:2Dautomatainclusions}.

%%%%%%

\section{Closure Properties}\label{sec:closure}

The closure properties of three-way two-dimensional automata are well-known. In the nondeterministic case, closure holds for the operations of union, row concatenation, 
and reversal (or row reflection) 
\cite{InoueTakanami1979ClosureProperties2DTuring, InoueTakanami1979TapeBounded2DTuring, Szepietowski1989ThreeWay2DTuringMachines, Szepietowski1992SomeRemarks2DAutomata}, while the model is not closed under the operations of intersection, complement, column concatenation, 
or rotation 
\cite{InoueTakanami1979TapeBounded2DTuring}. Focusing on the deterministic model, on the other hand, closure holds only for the operation of complement~\cite{Szepietowski1992SomeRemarks2DAutomata}.

For boustrophedon automata, it is known that the class of languages recognized by this model is closed under union, intersection, complement, row concatenation, and reversal~\cite{Fernau2015ScanningPicturesBoustrophedon, Fernau2018SimplePictureProcessing, Siromoney1973PictureLanguagesArrayRewriting}, but not under column concatenation or rotation~\cite{Fernau2018SimplePictureProcessing, Siromoney1973PictureLanguagesArrayRewriting}.

Returning to the two-way two-dimensional automaton model briefly mentioned earlier, it is known that the deterministic variant is closed under complement while the nondeterministic variant is closed under union; closure does not hold for any other operation~\cite{Smith2021PhDThesis, SmithSalomaa20212DConcatenation}.

Since typewriter automata are positioned between the two- and three-way two-dimensional automaton models in the recognition hierarchy, we can 
establish closure results for the typewriter automaton model 
using proofs analogous to those used for other models.

\begin{proposition}
$L_{\TYFA}$ is closed under union. $L_{\TYDFA}$ is closed under complement.
\end{proposition}

Things become more interesting, however, when we consider operations having positive closure results for the three-way model but not the two-way model; namely, the operations of row concatenation and reversal. Both operations are illustrated 
in Figure~\ref{fig:operations}, and the operations can be extended to two-dimensional languages in the usual manner. For row concatenation, we can make use of nondeterminism in a typewriter automaton to establish closure. First, however, we must make a slight modification to our model.

\begin{figure}
\centering
\begin{subfigure}[b]{0.49\textwidth}
\centering
\[\arraycolsep=1.4pt\def\arraystretch{0.8}
v \ominus w = 
\begin{array}{ccccc}
\#	& \#		&		& \#		& \# \\
\#	& v_{1,1}	& \cdots	& v_{1,n}	& \# \\
	& \vdots	& \ddots	& \vdots	& \\
\#	& v_{m,1}	& \cdots	& v_{m,n}	& \# \\
\#	& w_{1,1}	& \cdots	& w_{1,n}	& \# \\
	& \vdots	& \ddots	& \vdots	& \\
\#	& w_{m',1}& \cdots	& w_{m',n}& \# \\
\#	& \#		&		& \#		& \#
\end{array}
\]
\caption{Row concatenation of two words $v$ and $w$}
\label{fig:rowconcat}
\end{subfigure}
\hfill
\begin{subfigure}[b]{0.49\textwidth}
\centering
\[\arraycolsep=1.4pt\def\arraystretch{0.8}
w^{\text{R}} = 
\begin{array}{ccccc}
\#	& \#		&		& \#		& \# \\
\#	& w_{m,1}	& \cdots	& w_{m,n}	& \# \\
	& \vdots	& \ddots	& \vdots	& \\
\#	& w_{1,1}	& \cdots	& w_{1,n}	& \# \\
\#	& \#		&		& \#		& \#
\end{array}
\]
\caption{Reversal of a word $w$}
\label{fig:reversal}
\end{subfigure}
\caption{Illustrations of the operations of row concatenation and reversal}
\label{fig:operations}
\end{figure}

Let us say that a typewriter automaton is ``bottom-accepting" if it meets one of two conditions: either (i) upon following a transition to $q_{\text{accept}}$, the input head of the automaton repeatedly makes reset moves until it reaches the bottom border of its input word; or (ii) upon reading a boundary marker \# following a reset move, the automaton immediately halts and accepts if $q_{\text{accept}}$ is reachable from its current state. Note that, in the second condition, the boundary marker read following a reset move must necessarily be on the bottom border of the input word.

\begin{lemma}
Given a typewriter automaton $\mathcal{T}$, there exists an equivalent bottom-accepting typewriter automaton $\mathcal{T}'$.

\begin{proof}
We consider each of the two conditions for a typewriter automaton to be ``bottom-accepting" separately.

If $\mathcal{T}$ follows a transition to $q_{\text{accept}}$ at some point during its computation, then we may take $\mathcal{T}'$ to be the same as $\mathcal{T}$ apart from the existing transitions to $q_{\text{accept}}$. We add a new state $p$, redirect all existing transitions leading to $q_{\text{accept}}$ to instead lead to $p$, add a looping transition on $p$ to make a reset move upon reading $\Sigma \setminus \{\#\}$, and add a transition from $p$ to $q_{\text{accept}}$ upon reading a boundary marker \#.

If $\mathcal{T}$ reads a boundary marker \# following a reset move and transitions to some state $q$, then its input head can only read boundary markers for the remainder of its computation. We may then take $\mathcal{T}'$ to be the same as $\mathcal{T}$ apart from its transition upon reading the first boundary marker following a reset move. In this case, we check whether $q_{\text{accept}}$ is reachable from $q$ via some sequence of transitions on an arbitrary number of boundary markers, and we modify the transition function of $\mathcal{T}'$ to transition to $q_{\text{accept}}$ in the positive case or be undefined in the negative case.
\end{proof}
\end{lemma}

Using this modification, we present the closure result for row concatenation.

\begin{theorem}
$L_{\TYFA}$ is closed under row concatenation.

\begin{proof}
Let $L_{1}$ and $L_{2}$ be languages recognized by bottom-accepting nondeterministic typewriter automata $\mathcal{T}_{1}$ and $\mathcal{T}_{2}$, respectively. 
We construct a nondeterministic typewriter automaton $\mathcal{S}$ that recognizes the language $L_{1} \ominus L_{2}$ in the following way.

First, $\mathcal{S}$ simulates the bottom-accepting computation of $\mathcal{T}_{1}$. If this simulated computation accepts, then $\mathcal{S}$ nondeterministically makes some number of reset moves in the input word before simulating the bottom-accepting computation of $\mathcal{T}_{2}$ and accepting if $\mathcal{T}_{2}$ accepts. Since the row concatenation operation does not separate concatenated words by boundary markers, the nondeterministic reset moves allow $\mathcal{S}$ to guess when it has positioned its input head at the top-left symbol of the input word to $\mathcal{T}_{2}$.
\end{proof}
\end{theorem}

Moving to the reversal operation, we will again make use of the bottom-accepting version of our typewriter automaton model.

\begin{theorem}
$L_{\TYFA}$ is closed under reversal.

\begin{proof}
Let $L$ be a language recognized by a bottom-accepting nondeterministic typewriter automaton $\mathcal{T}$. We construct a nondeterministic typewriter automaton $\mathcal{T}'$ that recognizes the reversal language $L^{\text{R}}$ in the following way.

Since $\mathcal{T}$ is bottom-accepting, its input head will be positioned in the bottom border of the input word at the end of its accepting computation. At the beginning of the computation of $\mathcal{T}'$ on some word from the language $L^{\text{R}}$, its input head will be positioned at the top-left (non-boundary marker) symbol of the input word. For each row of the input word, $\mathcal{T}'$ searches the computation tree in reverse starting from a vertex $c_{\text{accept}}$ corresponding to an accepting configuration. If $\mathcal{T}'$ encounters a transition on the symbol currently being read by its input head at some vertex $c$ of the computation tree, then $\mathcal{T}'$ simulates the computation from vertex $c$ forward to vertex $c_{\text{accept}}$. When $\mathcal{T}'$ encounters the first transition resulting in a reset move during this simulation, it moves its input head to the next row and begins another reverse search starting from the vertex $c$. If, at some point, $\mathcal{T}'$ arrives at the initial vertex $c_{\text{init}}$ of the computation tree, then it checks for a boundary marker after making its first reset move during the simulated computation. If the input head reads a boundary marker after the reset move, then $\mathcal{T}'$ accepts.
\end{proof}
\end{theorem}

%%%

\subsection{Operational State Complexity}

Although state complexity is a well-studied measure for traditional automaton models, the same cannot be said for two-dimensional automaton models. The primary barrier to obtaining such results is the fact that there are currently no known general techniques for proving state complexity lower bounds in two dimensions, in contrast to the techniques known for one-dimensional (string) automata.

That being said, one-dimensional state complexity techniques have been used successfully to obtain bounds for projections of two-dimensional languages~\cite{SmithSalomaa20202DProjection}, and a number of elementary operational state complexity upper bounds have been obtained directly for certain two-dimensional language operations~\cite{Smith2021PhDThesis}. Of these elementary bounds, one pertaining to the three-way two-dimensional automaton model is particularly relevant:

\begin{proposition}\label{thm:2NFA3Wrowconcatstatecomplexity}
Given two nondeterministic three-way two-dimensional automata $\mathcal{A}$ and $\mathcal{B}$ with $m$ and $n$ states, respectively, the automaton $\mathcal{A} \ominus \mathcal{B}$ contains at most $m + n + 2$ states.
\end{proposition}

The idea underlying the proof of Proposition~\ref{thm:2NFA3Wrowconcatstatecomplexity} is that an additional two states are needed to move the input head from its accepting position in the input word to $\mathcal{A}$ to its initial position in the input word to $\mathcal{B}$.

Since the operation of row concatenation is closed for nondeterministic typewriter automata, we can investigate its state complexity. Due to the fact that typewriter automata use a single reset move to shift the input head leftward and downward, we immediately incur a savings in the number of states needed to recognize a row concatenation language.

\begin{theorem}
Given two nondeterministic typewriter automata $\mathcal{T}_{1}$ and $\mathcal{T}_{2}$ with $m$ and $n$ states, respectively, the automaton $\mathcal{T}_{1} \ominus \mathcal{T}_{2}$ contains at most $m + n$ states.

\begin{proof}
The proof of this theorem is analogous to the proof of Proposition~\ref{thm:2NFA3Wrowconcatstatecomplexity}, but with the two additional states---one for moving leftward repeatedly and one for moving downward to the following row---replaced by a single reset move transition.
\end{proof}
\end{theorem}

Note that $m + n$ states is presumably the matching lower bound for the row concatenation of two typewriter automaton languages, since the individual automata require $m$ and $n$ states to recognize their respective languages. However, as mentioned before, there is no known general technique for establishing lower bounds for two-dimensional automaton models.

%%%%%%

\section{Conclusion}\label{sec:conclusion}

In this paper, we introduced the notion of a typewriter automaton, drew connections between this model and the existing models of boustrophedon and returning automata, established recognition properties of the model relative to other two-dimensional automaton models, showed closure properties of certain language operations, and considered operational state complexity bounds. We showed that a typewriter automaton is capable of recognizing languages in between the two- and three-way variants of the standard two-dimensional automaton model, while the nondeterministic typewriter automaton has a recognition power equivalent to that of both boustrophedon and returning automata. We further established closure of the union, row concatenation, and reversal language operations for typewriter automata, and of the complement operation for the deterministic variant. Lastly, we investigated the operational state complexity of row concatenation for nondeterministic typewriter automata.

A number of fruitful questions remain surrounding the typewriter automaton model. For instance, knowing that the reversal operation is closed for nondeterministic typewriter automata and that the union operation is closed for the deterministic variant, it would be interesting to establish state complexity bounds for these operations, possibly using some as-yet-unknown proof technique. Beyond settling the question for typewriter automata, the same techniques may be applicable to boustrophedon automata or to the more general three-way two-dimensional automaton model. Future research may also consider the computational complexity of certain decision problems for typewriter automata, as has been done for boustrophedon automata.

In one of their papers on boustrophedon automata~\cite{Fernau2018SimplePictureProcessing}, Fernau et al.\ mention that boustrophedon scanning is not a new innovation, and that the technique is used widely in image processing applications; for example, in converting two-dimensional data to a one-dimensional string for compression, or in preprocessing data for optical character recognition. As the action of a typewriter automaton is somewhat similar to that of a boustrophedon (or a returning) automaton, it is not unreasonable to assume that the typewriter automaton model might also find applications in image processing or recognition. In particular, the left-to-right movement of the input head combined with access to the reset move may be suitable for applications such as edge finding or mapping contiguous segments in images represented as two-dimensional words. Alongside the restricted two- and three-way two-dimensional automaton models, typewriter automata present a conceptually simpler model of computation relative to the standard four-way two-dimensional automaton model.

%%%%%%

\section*{Acknowledgements}\label{sec:acknowledgements}

The author would like to thank Ian McQuillan for suggesting the idea of a ``two-dimensional typewriter automaton" during his PhD defence in August 2021.

%%%%%%

\bibliographystyle{plain}
\bibliography{./References.bib}

\begin{thebibliography}{10}

\bibitem{Fernau2015ScanningPicturesBoustrophedon}
Henning Fernau, Meenakshi Paramasivan, Markus~L. Schmid, and D.~Gnanaraj
  Thomas.
\newblock Scanning pictures the boustrophedon way.
\newblock In R.~P. Barneva, B.~B. Bhattacharya, and V.~E. Brimkov, editors,
  {\em Proceedings of the 17th International Workshop on Combinatorial Image
  Analysis ({IWCIA} 2015)}, volume 9448 of {\em Lecture Notes in Computer
  Science}, pages 202--216, Berlin Heidelberg, 2015. Springer-Verlag.

\bibitem{Fernau2018SimplePictureProcessing}
Henning Fernau, Meenakshi Paramasivan, Markus~L. Schmid, and D.~Gnanaraj
  Thomas.
\newblock Simple picture processing based on finite automata and regular
  grammars.
\newblock {\em Journal of Computer and System Sciences}, 95:232--258, 2018.

\bibitem{Fernau2016RegularArrayBoustrophedon}
Henning Fernau, Meenakshi Paramasivan, and D.~Gnanaraj Thomas.
\newblock Regular array grammars and boustrophedon finite automata.
\newblock In H.~Bordihn, R.~Freund, B.~Nagy, and G.~Vaszil, editors, {\em Short
  Papers of the 8th Workshop on Non-Classical Models of Automata and
  Applications ({NCMA} 2016)}, pages 55--63, Vienna, 2016. Institut f{\"u}r
  Computersprachen, TU Wien.

\bibitem{InoueTakanami1979ClosureProperties2DTuring}
Katsushi Inoue and Itsuo Takanami.
\newblock Closure properties of three-way and four-way tape-bounded
  two-dimensional {T}uring machines.
\newblock {\em Information Sciences}, 18(3):247--265, 1979.

\bibitem{InoueTakanami1979TapeBounded2DTuring}
Katsushi Inoue and Itsuo Takanami.
\newblock Three-way tape-bounded two-dimensional {T}uring machines.
\newblock {\em Information Sciences}, 17:195--220, 1979.

\bibitem{Inoue19912DAutomataSurvey}
Katsushi Inoue and Itsuo Takanami.
\newblock A survey of two-dimensional automata theory.
\newblock {\em Information Sciences}, 55(1--3):99--121, 1991.

\bibitem{KariSalo2011PictureWalkingAutomataSurvey}
Jarkko Kari and Ville Salo.
\newblock A survey on picture-walking automata.
\newblock In W.~Kuich and G.~Rahonis, editors, {\em Algebraic Foundations in
  Computer Science: Essays Dedicated to {S}ymeon {B}ozapalidis on the Occasion
  of His Retirement}, volume 7020 of {\em Lecture Notes in Computer Science},
  pages 183--213, Berlin Heidelberg, 2011. Springer-Verlag.

\bibitem{Pecuchet1985BoustrophedonInfinite}
{J}ean-{P}ierre P{\'e}cuchet.
\newblock Automates boustroph{\'e}don et mots infinis.
\newblock {\em Theoretical Computer Science}, 35:115--122, 1985.
\newblock In French.

\bibitem{Pecuchet1985BoustrophedonBirget}
{J}ean-{P}ierre P{\'e}cuchet.
\newblock Automates boustroph{\'e}don, semi-groupe de {B}irget et mono{\"\i}de
  inversif libre.
\newblock {\em {RAIRO} -- Theoretical Informatics and Applications},
  19(1):71--100, 1985.
\newblock In French.

\bibitem{Rosenfeld1979PictureLanguages}
Azriel Rosenfeld.
\newblock {\em Picture Languages: Formal Models for Picture Recognition}.
\newblock Computer Science and Applied Mathematics. Academic Press, New York,
  1979.

\bibitem{Siromoney1973PictureLanguagesArrayRewriting}
Gift Siromoney, Rani Siromoney, and Kamala Krithivasan.
\newblock Picture languages with array rewriting rules.
\newblock {\em Information and Control}, 22(5):447--470, 1973.

\bibitem{Smith2019TwoDimensionalAutomata}
Taylor~J. Smith.
\newblock Two-dimensional automata.
\newblock Technical report 2019-637, Queen's University, Kingston, 2019.

\bibitem{Smith2021PhDThesis}
Taylor~J. Smith.
\newblock {\em Closure, Decidability, and Complexity Results for Restricted
  Variants of Two-Dimensional Automata}.
\newblock {PhD} thesis, Queen's University, 2021.

\bibitem{SmithSalomaa20202DProjection}
Taylor~J. Smith and Kai Salomaa.
\newblock Recognition and complexity results for projection languages of
  two-dimensional automata.
\newblock In G.~Jir{\'a}skov{\'a} and G.~Pighizzini, editors, {\em Proceedings
  of the 22nd International Conference on Descriptional Complexity of Formal
  Systems ({DCFS} 2020)}, volume 12442 of {\em Lecture Notes in Computer
  Science}, pages 206--218, Berlin Heidelberg, 2020. Springer-Verlag.

\bibitem{SmithSalomaa20212DConcatenation}
Taylor~J. Smith and Kai Salomaa.
\newblock Concatenation operations and restricted variants of two-dimensional
  automata.
\newblock In T.~Bure{\v s} et~al., editors, {\em Proceedings of the 47th
  International Conference on Current Trends in Theory and Practice of Computer
  Science ({SOFSEM} 2021)}, volume 12607 of {\em Lecture Notes in Computer
  Science}, pages 147--158, Berlin Heidelberg, 2021. Springer-Verlag.

\bibitem{Szepietowski1989ThreeWay2DTuringMachines}
Andrzej Szepietowski.
\newblock On three-way two-dimensional {T}uring machines.
\newblock {\em Information Sciences}, 47(2):135--147, 1989.

\bibitem{Szepietowski1992SomeRemarks2DAutomata}
Andrzej Szepietowski.
\newblock Some remarks on two-dimensional finite automata.
\newblock {\em Information Sciences}, 63(1--2):183--189, 1992.

\end{thebibliography}
\nocite{*}

%%%%%%

\end{document}